\documentclass[useAMS,usenatbib]{mn2e}

\usepackage{graphicx}
\usepackage{natbib}
\usepackage[figuresleft]{rotating}
\newcommand{\HI}{H{\,\small I}}
\newcommand{\HII}{H{\,\small II}}
\newcommand{\Htwo}{H$_{2}$}
\newcommand{\ltsima} {$\; \buildrel < \over \sim \;$}
\newcommand{\gtsima} {$\; \buildrel > \over \sim \;$}
\newcommand{\lta} {\lower.5ex\hbox{\ltsima}}
\newcommand{\gta} {\lower.5ex\hbox{\gtsima}}

\newcommand{\kms}{km\ s$^{-1}$}

\newcommand{\mjybm}{mJy~bm$^{-1}$}
\newcommand{\lya}{Ly$\alpha$}
\newcommand{\CIV}{C{\ \small IV}}
\title[CO observations of two high-$z$ radio galaxies with CABB]{CO observations of high-$z$ radio galaxies MRC~2104-242 and MRC~0943-242: spectral-line performance of the Compact Array Broadband Backend}
\author[B. H. C. Emonts et al.]{B. H. C. Emonts$^{1}$\thanks{E-mail:bjorn.emonts@csiro.au}\thanks{Bolton Fellow}, R. P. Norris$^{1}$, I. Feain$^{1}$, G. Miley$^{2}$, E. M. Sadler$^{3}$,
\newauthor M. Villar-Mart\'{i}n$^{4}$, M. Y. Mao$^{5,6,1}$, T. A. Oosterloo$^{7,8}$, R. D. Ekers$^{1}$
\newauthor J. B. Stevens$^{1}$, M. H. Wieringa$^{1}$, K. E. K. Coppin$^{9}$ and C. N. Tadhunter$^{10}$\\
$^{1}$CSIRO Astronomy and Space Science, Australia Telescope National Facility, PO Box 76, Epping NSW, 1710, Australia\\
$^{2}$Leiden Observatory, University of Leiden, P.O. Box 9513, 2300 RA Leiden, Netherlands\\
$^{3}$School of Physics, University of Sydney, NSW 2006, Australia\\
$^{4}$Instituto de Astrof\'{i}sica de Andaluc\'{i}a (CSIC), Aptdo. 3004, Granada, Spain\\
$^{5}$School of Mathematics and Physics, University of Tasmania, Private Bag 37, Hobart, 7001, Australia\\
$^{6}$Australian Astronomical Observatory, PO Box 296, Epping, NSW, 1710, Australia\\
$^{7}$Netherlands Institute for Radio Astronomy, Postbus 2, 7990 AA Dwingeloo, the Netherlands\\
$^{8}$Kapteyn Astronomical Institute, University of Groningen, P.O. Box 800, 9700 AV Groningen, the Netherlands\\
$^{9}$Institute for Computational Cosmology, Durham University, South Road,Durham, DH1 3LE, UK\\
$^{10}$Department of Physics and Astronomy, University of Sheffield, Sheffield S3 7RH, UK\\
}
\begin{document}

\date{}

\pagerange{\pageref{firstpage}--\pageref{lastpage}} \pubyear{2010}

\maketitle

\label{firstpage}

\begin{abstract}
We present the first 7-millimetre observations of two high-redshift, Ly$\alpha$-bright radio galaxies (MRC~2104-242 and MRC~0943-242) performed with the $2 \times 2$ GHz instantaneous bandwidth of the Compact Array Broadband Backend (CABB) at the Australia Telescope Compact Array (ATCA). The aim was to search for $^{12}$CO(1-0) emission in these systems and test the millimetre capabilities of CABB for performing spectral line observations at high redshifts. We show that the stable band and enhanced velocity coverage of CABB, combined with hybrid array configurations, provide the ATCA with excellent 7-mm capabilities that allow reliable searches for the ground transition of CO at high redshifts. In this paper we explicitly discuss the calibration procedures used to reach our results. We set a firm upper limit to the mass of molecular gas in MRC~2104-242 ($z = 2.5$) of M$_{\rm H2} < 2 \times 10^{10}\ (\alpha_{\rm x}/0.8)$\,M$_{\odot}$. For MRC~0943-242 ($z=2.9$) we derive an upper limit of M$_{\rm H2} < 6 \times 10^{10}\ (\alpha_{\rm x}/0.8)$\,M$_{\odot}$. We also find a tentative 3$\sigma$ CO detection in the outer part of the giant \lya\ halo that surrounds MRC~0943-242. The 30-33 GHz radio continuum of MRC~2104-242 and MRC~0943-242 is reliably detected. Both radio sources show a spectral index of $\alpha \approx -1.5$ between 1.4 and 30 GHz, with no evidence for spectral curvature within this range of frequencies.

\end{abstract}


\section{Introduction}
\label{sec:intro}

Cold gas is a primary component in galaxy formation processes such as star formation and disk growth. However, despite detailed studies of cold gas in the nearby Universe, it is still difficult to trace similar quantities of cold gas beyond our Galactic backyard. Recently, \citet{tac10} and \citet{dad10} observed that star-forming galaxies at high redshifts are likely to contain a much larger fraction of their total mass in the form of molecular gas compared with nearby massive spiral galaxies. Recent simulations support this idea that the molecular gas content of galaxies increases when going to higher redshifts \citep{obr09_1,obr09_4,obr09_3,obr09_2}. These results demonstrate that extensive studies of cold molecular gas in the early Universe are becoming feasible with existing radio telescopes.

Powerful radio galaxies enable comprehensive studies of the cold ISM throughout the Universe. Their strong radio sources provide a background continuum against which we can search for foreground neutral and molecular gas in absorption \citep[e.g.][]{uso91,vermeu03,kan07,car07}, while their host galaxies are generally in a very specific stage of galaxy evolution. Detailed studies at low and intermediate redshifts reveal that powerful radio galaxies are frequently associated with gas-rich galaxy mergers \citep[e.g.][]{hec86,bau92}, often contain young stellar populations \citep{tad05,hol07,lab08}, and many display strong jet-ISM interactions \citep{tad91,vil99,cla98,emo05,mor05,mor05_HI,hol08}. At high redshifts ($z > 2$), luminous radio galaxies ($L_{\rm 500 MHz} > 10^{27}$ W~Hz$^{-1}$) are among the most massive galaxies in the early Universe \citep[see][for a review]{mil08}. They are typically surrounded by proto-clusters, which are thought to be the ancestors of rich local clusters \citep[e.g.][]{pen00,ven07}. The high-$z$ radio galaxies and surrounding proto-cluster gas and galaxies often interact with one another \citep[e.g.][]{nes08,ivi08} and are therefore laboratories for studying the formation and evolution of galaxies and clusters as well as investigating the relationship between early star formation and AGN activity. 

Since \citet{bro91} first observed CO gas (the strongest tracer for molecular hydrogen) at a redshift beyond $z=2$, intensive searches for CO in high-$z$ radio galaxies during the early 1990s were unsuccessful \citep{eva96,oji97}. Since then, studies of individual radio galaxies at $z \sim 2-5$ with synthesis radio telescopes have found CO emission (tracing molecular gas masses of a few $\times 10^{10}-10^{11} M_{\odot}$) in a number of these systems \citep[e.g.][see also \citet{sol05,omo07,mil08} for reviews]{sco97,pap00,bre03,bre03AR,bre05,kla05,nes09}. In some cases CO is observed to be resolved on scales of several tens of kpc \citep[e.g.][]{pap00}. This indicates that large amounts of cold molecular gas may be relatively common in high-$z$ radio galaxies. However, the major observational limitations for starting comprehensive studies of CO in high-$z$ radio galaxies have been the very limited velocity coverage of existing mm-spectrometers (often not much wider than the velocity range of the CO gas and/or the accuracy of the redshift) plus the fact that most observatories can only target the higher order rotational transitions of $^{12}$CO. 

Although the higher order CO lines are likely to have a higher flux density than the lower ones in the nuclear starburst/AGN regions, where gas is dense and thermally excited, \citet{pap00,pap01} suggest that the opposite may be true for large reservoirs of less dense and sub-thermally excited gas that is more widely distributed. In fact, various studies of the low-order CO transitions in different types of high-$z$ galaxies reveal molecular gas that is sub-thermally excited\footnote{\citet{har10} and \citet[][]{dan10} point out that a multi-component inter-stellar medium, rather than sub-thermal excitation, may better reflect the physical properties of the molecular gas in high-$z$ systems.} \citep{gre03,hai06,dan09,rie10} or distributed in extended reservoirs \citep{dad10,car10,ivi10,ivi11}. Cold CO gas distributed across the host galaxy may thus be much easier to detect in the lower CO transitions than generally assumed from studies of the higher transitions. Moreover, with uncertainties in excitation properties of the gas, observations of the rotational ground-transition of the CO molecule [$^{12}$CO(1-0) -- referred to as CO in the remainder of this paper] provide the most accurate mass estimate of the overall molecular gas content in these systems.

Since April 2009, the Australia Telescope Compact Array (ATCA) has a new broad-band backend system (the Compact Array Broadband Backend or CABB). CABB offers an instantaneous bandwidth of 4 GHz, split over $2 \times 2$ GHz observing bands, both with all Stokes polarisation parameters and 2048 channels (i.e. spectral resolution of 1 MHz); see \citet{fer02,wil11}. ATCA/CABB has millimetre observing capabilities at 3mm ($83.9-104.8$ GHz), 7mm ($30.0-50.0$ GHz) and 15mm ($16.0-25.0$ GHz). This, in combination with hybrid array configurations with baselines as short as 31m, makes the upgraded ATCA an excellent facility to detect and spatially resolve molecular gas in high-$z$ radio galaxies by targeting the lower rotational CO transitions (see Sect. \ref{sec:observations} for more details). A remarkable example of this is the recent detection of CO(2-1) in the distant ($z = 4.8$) sub-millimetre galaxy LESS~J033229.4-275619 by \citet{cop10}.

To test the spectral-line performance of CABB over the $2 \times 2$~GHz bandwidth, we used the 7mm band to search for CO(1-0) in two high-$z$ radio galaxies from the Molonglo Reference Catalogue \citep{mcc90}, namely MRC~2104-242 ($z=2.5$) and MRC~0943-242 ($z=2.9$). These two sources are part of a larger sample of high-$z$ radio galaxies that we aim to target with CABB in order to perform a systematic search for CO(1-0) in these systems.

MRC~2104-242 and MRC~0943-242 both have a redshift that corresponds to a critical epoch in galaxy formation ($z \sim 2.5-3$), at which there is a dramatic increase in sub-mm flux \citep{arc01,sma02,cha05} and the space-density of (radio-loud) quasars reaches a maximum \citep[e.g.][]{pei95,sha96,ric06}. {\it HST} observations by \citet{pen01} show that  MRC~2104-242 and MRC~0943-242 both have an optical continuum that is clumpy and elongated in the direction of the radio source \citep{pen00_radio,car97}. \citet{vil03} show that they both contain a giant Ly$\alpha$-halo ($\geq$ 100 kpc in diameter). For MRC~2104-242 the Ly$\alpha$ gas is distributed roughly along the radio axis in what appears to be a rotating structure with a diameter $\ga 120$ kpc \citep{vil06}. MRC~0943-242 shows a quiescent \lya-halo that extends well beyond the radio structure \citep{vil03}. MRC~0943-242 also shows a deep Ly$\alpha$ absorption \citep{rot95,jar03}, indicating that large amounts of neutral gas are present in this system. From fitting the spectral energy distributions of the host galaxies with {\it Spitzer}, \citet{sey07} derive a total stellar mass of a few $\times 10^{11} M_{\odot}$ for both systems.

In Sect. \ref{sec:observations} we present our CO observations and explain in more detail the enhanced capabilities of the ATCA for studying molecular gas at high redshifts. Section \ref{sec:results} shows the result regarding both the performance of CABB for doing these high-$z$ CO studies as well as the scientific outcome of our observations of MRC~2104-242 and MRC~0943-242. In Sect. \ref{sec:discussion} and \ref{sec:conclusions} we discuss the scientific results and conclude that the upgraded ATCA is a world-class facility for spectral line observations of the cold molecular gas at high-$z$.

\section{Observations}
\label{sec:observations}

During the period May - September 2009, MRC~2104-242 and MRC~0943-242 were observed with ATCA/CABB. Details of the observations are given in Table \ref{tab:observations}. 

\begin{figure}
\centering
\includegraphics[width=0.47\textwidth]{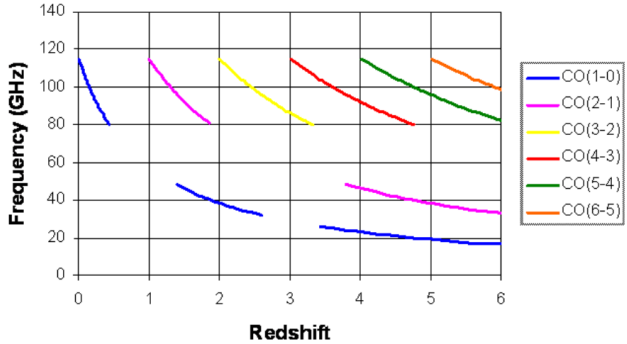}
\caption{Observing frequency of the first six rotational transitions of CO that can be targeted with the CABB millimetre system (3, 7 and 15mm) plotted against the redshift of the CO-emitters.}
\label{fig:ATCAbands}
\end{figure}

Figure \ref{fig:ATCAbands} shows the observing windows for the various transitions of extra-galactic CO currently available with CABB in the 3, 7 and 15 millimetre bands. 
For MRC~2104-242 and MRC~0943-242 we targeted the ground-transition CO(1-0) with the CABB 7mm system. The redshift of MRC~2104-242 ($z=2.491$) corresponds to an observing frequency of 33.0 GHz, which is also one of the optimum CABB frequencies for centring the 7mm band.\footnote{See the online ATCA Users Guide for details: http://www. narrabri.atnf.csiro.au/observing/users$\_$guide/html/atug.html} The redshift of MRC~0943-242 ($z=2.9185$)\footnote{This redshift corresponds to the velocity of the most prominent \HI\ absorption in the Ly$\alpha$ profile of MRC~0943-242 \citep{jar03}; see Sect. \ref{sec:fluxcal} for more details.} corresponds to an observing frequency of 29.4 GHz, which is outside the nominal 7mm CABB band ($30.0-50.0$ GHz). Nevertheless, when centring the band at 30.001 GHz, data is obtained down to 29 GHz. Observations of MRC~0943-242 therefore served as a good test of how well the CABB system behaves at the very edge of the 7mm band.

\begin{figure*}
\centering
\includegraphics[width=\textwidth]{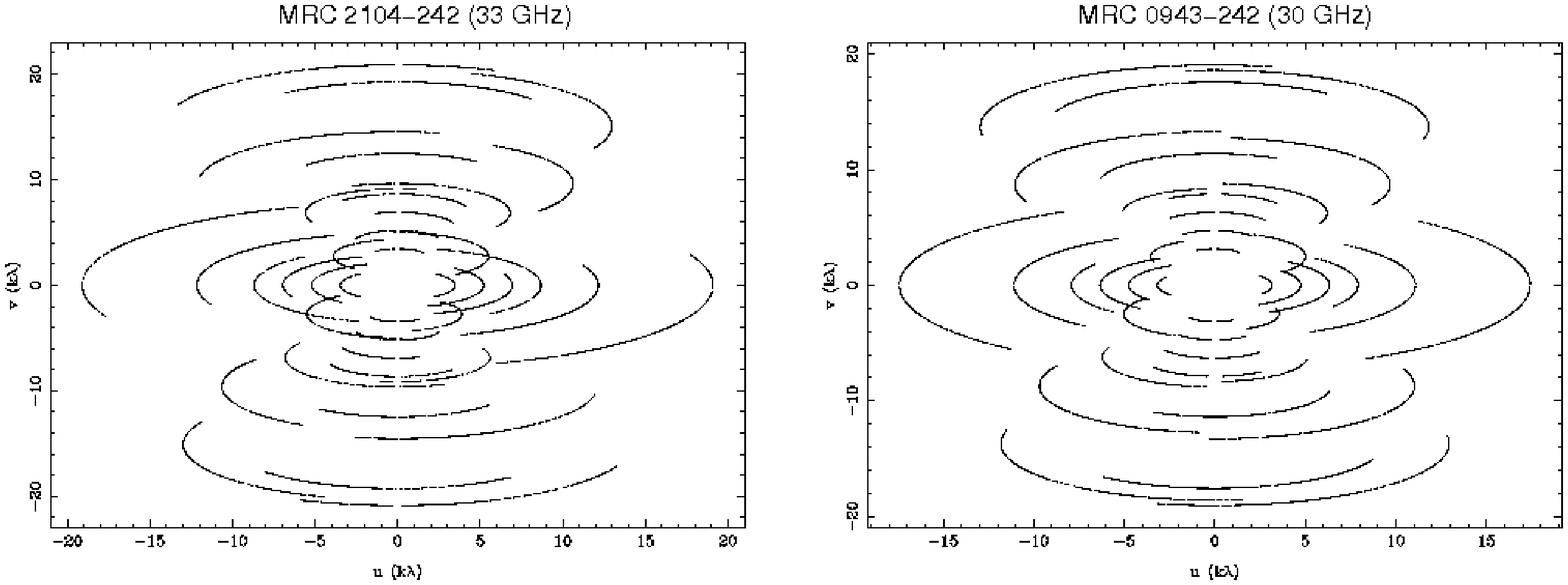}
\caption{The {\it uv}-coverage of the combined observing runs of both MRC~2104-242 and MRC~0943-242 (Table \ref{tab:observations}), using the hybrid H75 and H168 arrays.}
\label{fig:uvcoverage}
\end{figure*}

The total observing time for each source -- including overhead and calibration -- was roughly 40 hours (see Table \ref{tab:observations} and Sect. \ref{sec:calibration}). The observations were spread over the two most compact hybrid array configurations (H75 and H168) in order to minimise the effect of atmospheric phase fluctuations \citep[which worsen with increase in baseline length;][]{kla_thesis}. Both array configurations include five antennas that are spread across both an east-west as well as a north-south spur. This ensures a reasonably good {\it  uv}-coverage (see Fig.\ref{fig:uvcoverage}) even from observing runs as short as 6-8h, during which a source is targeted only above an elevation of 30$^{\circ}$ \citep[in order to avoid high airmasses, which dramatically increase the system temperature; see][]{kla_thesis}. The system temperatures ranged from $70-132$K for the observations of MRC~2104-242 and from $130-180$K for MRC~0943-242.\footnote{More details on theoretical estimates of T$_{\rm sys}$ values at 7mm can be found in the online ATCA Users Guide.}

\begin{table}
\centering
\caption{Observations}
\vspace{0.4cm}
\label{tab:observations}
\begin{tabular}{lllccc}
Source & Array & Obs. date & t$_{\rm int}$ (h) & $\nu_{\rm central}$ (GHz) \\
\hline
MRC~2104-242 & H75  & 09JUL21  & 6.99 & 33.000 \\
             &      & 09JUL23  & 4.15 & 33.020 \\
             & H168 & 09SEP18  & 3.00 & 33.000 \\
             &      & 09SEP19  & 3.14 & 33.000 \\
             &      & 09SEP20  & 2.19 & 33.000 \\
MRC~0943-242 & H75  & 09JUL11  & 3.10 & 30.001 \\
             &      & 09JUL12  & 3.20 & 30.010 \\
             &      & 09JUL14  & 2.88 & 30.001 \\
             & H168 & 09MAY10  & 3.29 & 30.001 \\
             &      & 09MAY11  & 2.68 & 30.015 \\
             &      & 09MAY12  & 3.18 & 30.001 \\
\end{tabular} 
\flushleft 
{Notes -- t$_{\rm int}$ is the effective on-source integration time (i.e. not including overheads).}
\end{table} 

At the high central frequencies of our observations (30-33 GHz), the coarsest spectral-line mode of CABB ($2 \times 2$~GHz bands with 1\,MHz spectral resolution) provides a velocity resolution of 9.5-11 \kms\ across an effective velocity coverage of at least 17,000 \kms\ per 2~GHz band. Both 2~GHz bands were centred around the same observing frequency, but because they are not mutually independent, only one 2~GHz band was used in the final data analysis.

Observations were done as much as possible during the night and in good weather conditions (to avoid decorrelation due to atmospheric phase instabilities) and only above 30$^{\circ}$ elevation (to avoid high system temperatures due to large airmasses). Because our observations also served as a test for the performance of CABB for spectral-line observations, in the following Section we will explain the details of several crucial calibration steps. For the data reduction and analysis we used {\sc MIRIAD} and {\sc KARMA}.

\subsection{Calibration, overheads and data reduction}
\label{sec:calibration}
Our general observing strategy was as follows: a strong calibrator was observed at least three times during each run in order to check the reliability of the bandpass calibration. A secondary (phase/gain) calibrator was observed roughly every 10 minutes. Flux calibration was done at least once during each run. Pointing solutions of the antennas were checked and updated every hour, or every time the telescope slewed more than $\sim$20$^{\circ}$ on the sky. Taking into consideration the conservative nature of this calibration strategy, the overheads due to calibration and slewing were about 50$\%$.\footnote{We estimate that in order to reach the potential maximum efficiency with less conservative calibration, overheads should be considered to be at least 30$\%$.}

\subsubsection{Phase/gain calibration}
\noindent For phase calibration we performed a 2 minute scan on a calibrator close to our target source roughly every 10 minutes, although target scans were decreased to 5 minutes in poor weather conditions and increased to 15 minutes when atmospheric phase stability was excellent. For MRC~2104-242 we used PKS~B2008-159, PKS~B2128-123 or PKS~2149-306 as phase calibrator. For MRC~0943-242 we used PKS~0919-260. Phase calibration was done in a standard way.

\subsubsection{Bandpass calibration}
\noindent In order to test the quality of the bandpass calibration at 7mm across the full 2 GHz band, we observed a strong calibrator (PKS~B0537-441, PKS~B1253-055, PKS~B1334-127, PKS~B1921-293 or PKS~B2223-052) at least three times during each run (unless the run was cut short due to weather). We noticed that weather and atmospheric conditions at the ATCA site can introduce frequency dependent temporal gain fluctuations across the wide CABB band, which can have a significant effect on the quality of the bandpass calibration at 7mm. It is therefore essential to obtain at least one good scan on the bandpass calibrator during good atmospheric conditions. For MRC~0943-242 we chose the best quality bandpass calibrator scan for calibrating our data. In case more than one bandpass calibrator scan was deemed suitable, we applied the bandpass solutions to that part of the data observed closest in time to the respective calibrator. 

For MRC~2104-242 the strong phase calibrators PKS~B2008-159 and PKS~B2128-123  (with observed fluxes of F$_{\rm 33~GHz} \approx 1.9$ and 1.8 Jy respectively) were suitable for bandpass calibration. This allowed us to obtain a bandpass solution roughly every 10-15 minutes. We used a new feature in the {\sc MIRIAD} task {\sc mfcal} to interpolate between consecutive bandpass solutions in order to compensate for possible frequency dependent gain variations that slowly fluctuate in time.

\subsubsection{Flux calibration}
\label{sec:fluxcal}
\noindent For MRC~2104-242, flux calibration was done by observing Uranus at the time that it was at roughly the same elevation as the phase calibrator and target source during each run. The presence of a weak radio continuum from the lobe-dominated high-$z$ radio galaxies in our 7mm data (which are not expected to significantly change their flux densities over time-scales of a few months) allowed us to compare the relative flux calibration between the various runs, which remained constant within 15$\%$. Our absolute flux calibration used the available {\sc MIRIAD}-model for Uranus. This model did not take into account changes in the planet's orientation, which introduce time-variations of up to 10$\%$ in its brightness temperature \citep[see][]{kra08,wei10}, potentially leading to a significant error in absolute flux calibration. During one of the runs we also observed PKS~B1934-638, which confirmed our Uranus-based absolute flux calibration to an accuracy of $\sim$18$\%$. We therefore estimate the overall (relative + absolute) uncertainty in the flux calibration of MRC~2104-242 to be within 30$\%$.

For MRC~0943-242, Uranus was not visible during our observing runs. For flux calibration we therefore observed the ultra-compact \HII\ region G309 [G309.9206+00.4790; \citet{urq07}, with our pointing centred at RA(J2000)=13:50:42.35, dec(J2000)=-61:35:09.78] when it was at roughly the same elevation as the phase calibrator. We calibrated the flux of G309 against Uranus, which we observed roughly half a day later for each run. The flux of G309 was stable over our six observing epochs and the relative flux calibration between the six different runs was within 13$\%$. From our data we derive a value of $S_{\rm 30~GHz} = 1.31 \pm 0.07$\,Jy for the shortest baselines at which the source is unresolved. Recently, \citet{mur10} derived a flux density of $S_{\rm 32~GHz} = 1.1 \pm 0.11$\,Jy for G309, also using Uranus as flux calibrator. In order to verify the accuracy of our absolute flux calibration, we observed PKS~B1934-638 during three of our observing epochs. When using PKS~B1934-638 as flux calibrator instead of Uranus, the absolute fluxes derived from our data are on average $\sim$\,15\,$\%$ lower. This uncertainty in absolute flux calibration is consistent with the difference between our flux estimate for G309 (which we used to calibrate our data) and that made by \citet{mur10}. This may again reflects variations in the brightness of Uranus that were not accounted for by the existing models (see previous paragraph). The spectral index of G309 changes at most a few percent across the 2~GHz band at 30 GHz, in agreement with \citet{mur10}. In all, we therefore estimate that for MRC~0943-242 the overall (relative + absolute) uncertainty in our flux calibration is within 30$\%$.\\
\vspace{0mm}\\
After flagging and bandpass, gain and flux calibration, we subtracted the continuum from the line data in the {\it uv}-domain by applying a linear fit to the channels across the full 2 GHz band (for MRC~0943-242 we excluded from this fit the channels in which we found a tentative CO signal, see Sect. \ref{sec:0943}, although this has no significant effect when fitting the full 2~GHz band). Subsequently, a robust +1 weighted \citep{bri95} continuum map and line data set were created by Fourier transforming the {\it uv}-data and, in case of the continuum map, cleaning the signal. We then translated the velocity axis to match the optical, barycentric rest-frame velocity at the redshift of MRC~2104-242 and MRC~0943-242. The redshift of MRC~2104-242 ($z = 2.491$) has been confirmed by \citet{ove01} and \citet{vil03} through observations of the \lya\ and various metal emission-lines. For MRC~0943-242, we chose to centre our observations at the redshift of the prominent \HI\ absorption in the Ly$\alpha$ profile \citep[$z=2.9185$;][]{jar03}, which likely represents the bulk of the cold neutral gas in this system \citep[see also][]{bre03AR}. Table \ref{tab:data} shows details of the final data products that we obtained from our observations (some of these are described further in Sect. \ref{sec:results}).

\begin{table}
\centering
\caption{Data}
\vspace{0.4cm}
\label{tab:data}
\begin{tabular}{lcc}
      & MRC~2104-242 & MRC~0943-242 \\
\hline
Effective int. time (h) & 19.5 & 18.3 \\
Target frequency (GHz) & 33.02 & 29.417 \\
Redshift & 2.491 & 2.9185 \\
Beam size (arcsec$\times$arcsec) & 11.7 $\times$ 7.6 & 11.5 $\times$ 9.0 \\
Beam PA [PA ($^{\circ}$)] & 97.2 & 87.5 \\
$\Delta$v (\kms) & 9.6 & 11.0 \\
$\sigma$$_{\rm cont}$ ($\mu$Jy bm$^{-1}$) & 29 & 33 \\
$S_{\rm cont}$ (mJy) & 4.0 & 3.3 \\
$\sigma$$_{\rm line}$ (\mjybm ch$^{-1}$) & 0.45 & 0.90 \\
$L'_{\rm CO}$ (K \kms pc$^{2}$)  & $< 2.6 \times 10^{10}$ & $< 7.3 \times 10^{10}$ \\
\end{tabular} 
\flushleft 
{Notes -- Effective int. time is the total effective on-source integration time of all runs in both configurations combined. Target frequency (GHz) is the observing frequency of the expected CO(1-0) line at the redshift of our sources (see text for details). $\Delta$v (\kms) is the velocity resolution per 1~MHz channel. $\sigma_{\rm cont}$ is the rms noise level of the continuum image after t$_{\rm int}$. $S_{\rm cont}$ is the integrated continuum flux of the radio source at the target frequency. $\sigma_{\rm line}$ is the rms noise level of the full-resolution line data per 1~MHz channel after t$_{\rm int}$. $L'_{\rm CO}$ gives the upper limit on the CO luminosity (see text for details).}
\end{table}

\section{Results}
\label{sec:results}

\subsection{CABB performance}
\label{sec:CABBperformance}

Figure \ref{fig:MRC2104} shows the 33~GHz radio continuum map of MRC~2104-242 and a spectrum at the location of the centre of the host galaxy. The continuum image has an rms noise level of $29$ $\mu$Jy beam$^{-1}$ (after t$_{\rm int}$ = 19.5h; see Table \ref{tab:data}), demonstrating the effectiveness of ATCA/CABB for deep millimetre continuum studies. The 2\,GHz spectrum has a large velocity coverage  of $\sim 17,000$ \kms\ with an rms noise in each 1~MHz channel ($\Delta$v = 9.6 \kms) of $\sigma = 0.45$ \mjybm, with no significant systematic bandpass effects. 

\begin{figure*}
\centering
\includegraphics[width=\textwidth]{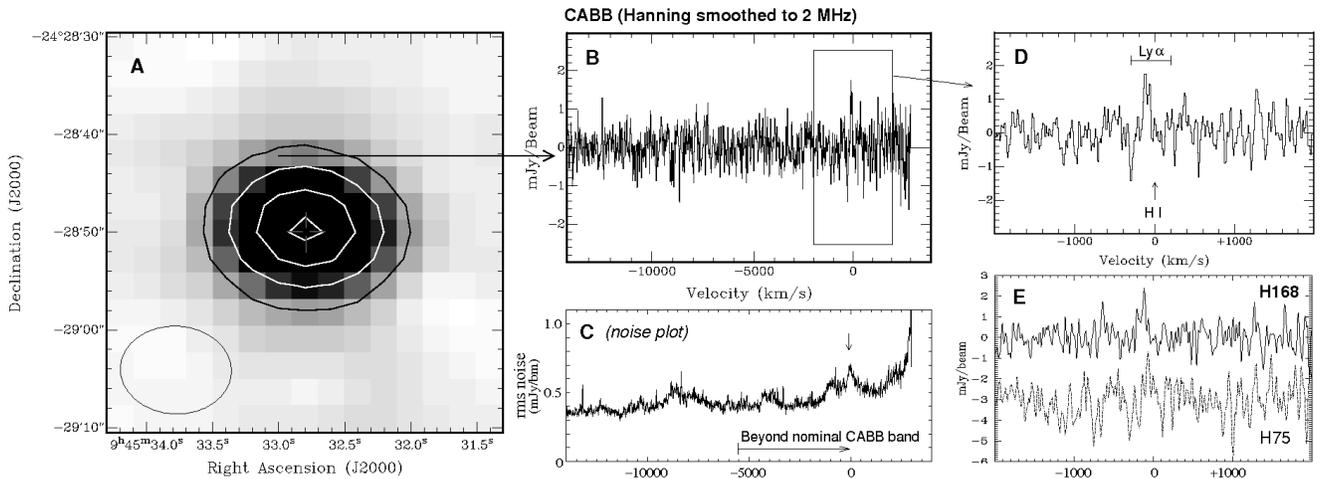}
\caption{{\sl a).} 30~GHz radio continuum map of MRC~0943-242 (contour levels: 0.4, 1.0, 2.0, 3.0 \mjybm). The cross indicates the location of the radio host galaxy. {\sl b).} Off-nuclear spectral line profile of redshifted CO. The spectrum is taken at the location marked by the arrow in figure {\sl a)} and Hanning smoothed to a velocity resolution of 2~MHz across the 2 GHz CABB band. The x-axis shows the optical barycentric velocity in the rest-frame of the radio host galaxy (see Sect \ref{sec:fluxcal}). {\sl c).} rms noise per channel in the Hanning smoothed data of figure {\sl b)} across the CABB band, derived across the central region. The arrow marks the rms noise level at the velocity of the tentative CO detection (right plot). {\sl d).} Zoom-in of figure {\sl b}, showing the tentative off-nuclear CO detection. The arrow indicates the redshift of the deep \lya\ absorption of \HI\ gas \citep{jar03} at which we centred our zero-velocity. The range of velocities of the emission-line gas in the giant \lya\ halo \citep{vil03} is also indicated in the plot. {\sl e).} separate data-sets of the H75 and H168 array observations, both showing the tentative CO signal (for illustration purposes, the x-axis of the H75-array data is scaled-down by 3 \mjybm\ in this plot).}
\label{fig:MRC0943}
\end{figure*}

Figure \ref{fig:MRC0943} shows the 30~GHz radio continuum map of MRC~0943-242 (with an rms noise of 33 $\mu$Jy beam$^{-1}$) and an off-nuclear spectral line profile. In this case, at the edge of the 7mm band, half the observing band lies outside the nominal CABB range (Sect. \ref{sec:observations}), where there are instrumental low-level structures in the noise or in the bandpass at about the 1$\sigma$ level of the full (1\,MHz) resolution data (Fig. \ref{fig:MRC0943}{\sl b}). In addition, the noise starts to vary beyond the nominal CABB range (Fig. \ref{fig:MRC0943}{\sl c}). After an effective on-source integration time of 18.3h, we derive a noise level at 29.4~GHz of $\sigma = 0.9$ \mjybm\ per 1~MHz channel ($\Delta$v = 11 \kms), i.e. twice the noise level at the optimum observing frequency of 33~GHz (see above). However, as can be seen in the Hanning-smoothed data of Fig. \ref{fig:MRC0943}{\sl c}, the noise level peaks at our target frequency of 29.4~GHz and is significantly lower throughout most part of the band, even below the nominal edge of 30 GHz. We therefore conclude that up to $\sim 0.8$ GHz below the nominal 7mm band, CABB is still suitable for spectral-line work.

\citet{cop10} detected CO(2-1) in a $z = 4.8$ sub-millimetre galaxy, which was observed with CABB at 40.0 GHz (i.e. towards the other end of the 7mm band compared to our 33/30 GHz observations). They find noise levels of $\sigma \approx 0.44$ \mjybm\ per 1 MHz channel and a bandpass stable enough to detect their CO signal at about the 5$\sigma$ level when binning across $>$10 channels. The data quality at 40~GHz thus appears comparable to that at 33~GHz as presented in this paper, giving a good indication for the excellent performance of CABB across the entire ATCA 7mm band.

\subsection{MRC~2104-242}
\label{sec:2104}

\begin{figure*}
\centering
\includegraphics[width=\textwidth]{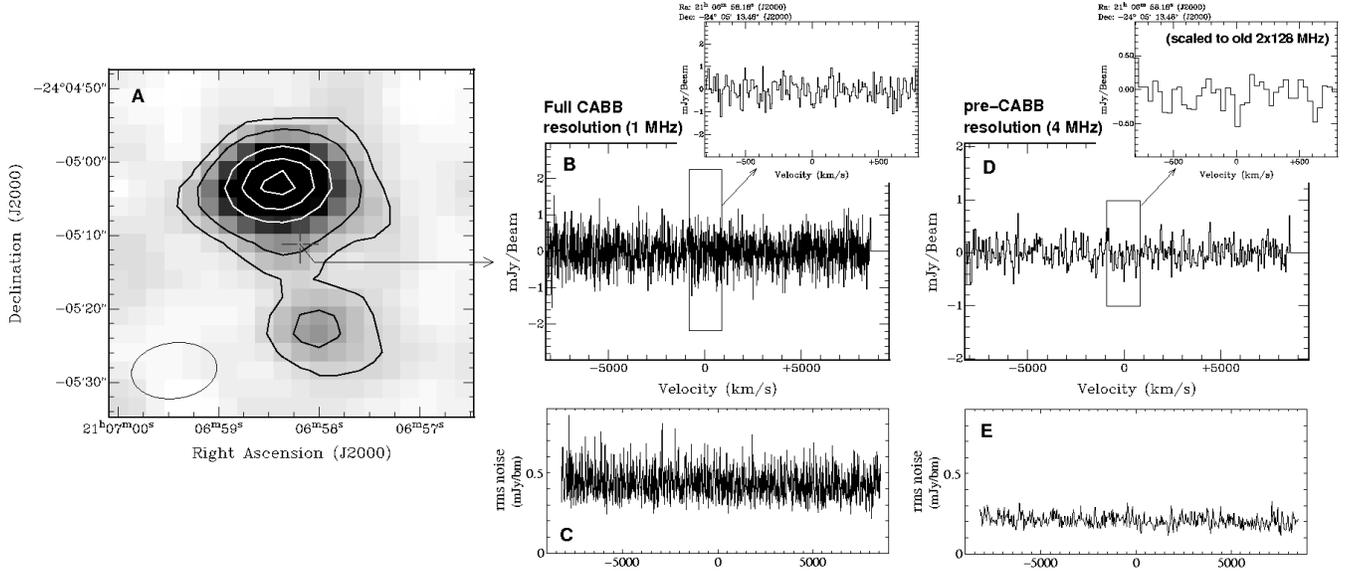}
\caption{{\sl a).} 33~GHz radio continuum map of MRC~2104-242 (contour levels: 0.1, 0.3, 1.0, 1.7, 2.4 \mjybm). The cross indicates the location of the radio host galaxy. {\sl b).} spectral line profile against the centre of the radio host galaxy. Shown is the full 1 MHz velocity resolution of CABB across the 2 GHz band. The x-axis shows the velocity in the rest-frame of the radio host galaxy (see Sect \ref{sec:fluxcal}). The zoom-in shows a portion of the CABB data with the approximate bandwidth coverage of the old pre-CABB ATCA system ($2 \times 128$ MHz). {\sl c).} rms noise per 1~MHz channel in the region of the radio source. {\sl d).} Same as figure {\sl b}, but data binned to 4~MHz channels (i.e. similar to the pre-CABB system; the zoom-in therefore gives a good representation of the data that could be obtained with the old $2\times 128$ MHz pre-CABB backend). {\sl e).} rms noise per channel of 4~MHz in the region of the radio source.}
\label{fig:MRC2104}
\end{figure*}

\begin{figure}
\centering
\includegraphics[width=0.4\textwidth]{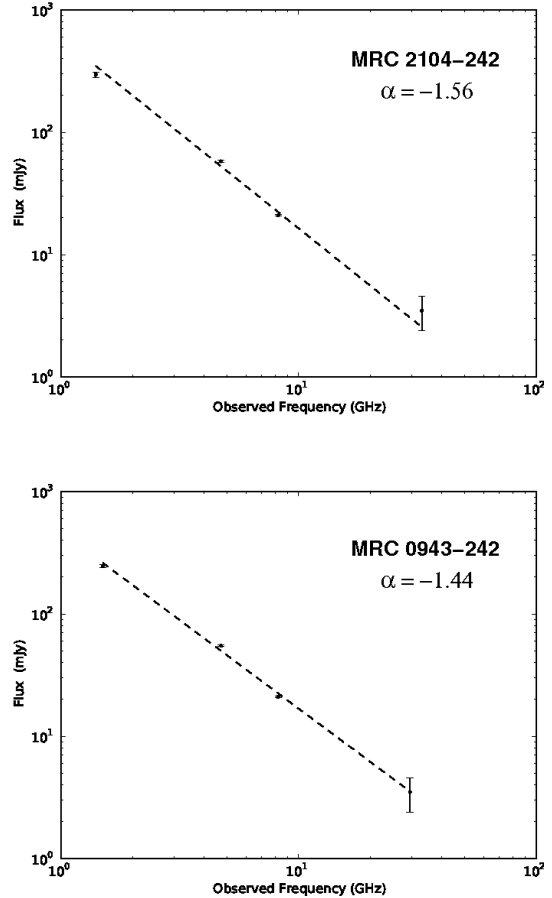}
\caption{Spectral index of MRC~2104-242 {\sl (top)} and MRC~0943-242 {\sl (bottom)}. Shown is the total integrated flux. Data at 1.4, 4.7 and 8.2 GHz are taken from \citet{car97} and \citet{pen00_radio}, with errors corresponding to 2$\%$ of the flux at these wavelengths \citep[as estimated by][]{car97}.}
\label{fig:index}
\end{figure}

MRC~2104-242 (Fig. \ref{fig:MRC2104}{\sl a}) is resolved at 33\,GHz with a total flux of 3.5 mJy. The continuum structure consists of two components on either side of the optical host galaxy, in agreement with 4.7 and 8.2 GHz VLA continuum observations that identified it as a double lobed radio source \citep{pen00_radio}. The bright northern lobe has a peak flux density of $S_{\rm 33 GHz} = 2.7$ \mjybm, while the fainter southern lobe has $S_{\rm 33 GHz} = 0.39$ \mjybm. Even at 33 GHz the radio continuum structure is dominated by the radio lobes and no core component (at the location of the optical nucleus) is seen in our data. We set a conservative upper limit to the 33 GHz core flux density of $S_{\rm core - 33 GHz} < $ 0.4 \mjybm. Figure \ref{fig:index} shows that the integrated flux of MRC~2104-242 has a steep spectral index between 1.4 GHz and 33.0 GHz, with $\alpha = -1.56$ (where $F_{\nu} \propto {\nu}^{\alpha}$). There is no evidence for spectral curvature within this range of frequencies. This is in agreement with spectral index observations of high-$z$ ultra-steep spectrum radio sources by \citet{kla06}.

No CO is detected in MRC~2104-242, either at the location of the host galaxy or at the position of the radio source. We derive a firm upper limit on the CO emission-line luminosity in MRC~2104-242 by assuming a potential 3$\sigma$ signal smoothed across 500 \kms, using
\begin{equation}
S_{\rm CO}\Delta {\rm V} = 3\sigma \ \Delta {\rm v} \ \sqrt{\frac{500\ {\rm km}\ {\rm s}^{-1}}{\Delta {\rm v}}}\ {\rm Jy} \cdot {\rm km}\ {\rm s}^{-1},
\label{eq:sco}
\end{equation}
with $\sigma$ the noise level per 1~MHz channel in one beam (in Jy) and $\Delta$v the width of one 1~MHz channel (in \kms). The CO luminosity (upper limit) can then be calculated following \citet[][and references therein]{sol05}:
\begin{equation}
L'_{\rm CO} = 3.25 \times 10^7 (\frac{S_{\rm CO} \Delta {\rm V}}{{\rm Jy}~{\rm km/s}}) (\frac{D_{\rm L}}{{\rm Mpc}})^2 (\frac{{\nu_{\rm rest}}}{\rm GHz})^{-2} (1+z)^{-1},
\label{eq:lco}
\end{equation}
with $L'_{\rm CO}$ expressed in ${\rm K~km/s~pc^2}$ and with $D_{\rm L} = 20018$ Mpc the luminosity distance of MRC~2104-242 \citep[following][]{wri06}\footnote{See http://www.astro.ucla.edu/$\sim$wright/CosmoCalc.html for Ned Wright's online cosmology calculator that we used to deriving luminosity and angular-size distances. Throughout this paper we use H$_{0} =71~{\rm km s}^{-1}~{\rm Mpc}^{-1}$, $\Omega_{\rm M} = 0.3$ and $\Omega_{\Lambda} = 0.7$.}. For MRC~2104-242, $S_{\rm CO}\Delta {\rm V} < 0.094$ Jy \kms, hence $L'_{\rm CO} < 2.6 \times 10^{10}\ {\rm K~km~s^{-1}~pc^2}$.

\subsection{MRC~0943-242}
\label{sec:0943}

The radio source MRC~0943-242 (Fig. \ref{fig:MRC0943}{\sl a}) has a flux density of 3.3 \mjybm\ and is unresolved in our data. Higher resolution continuum observations at 4.7 and 8.2 GHz by \citet{car97} show that the radio source consists of two lobes that are separated by 4 arcsec. When comparing the flux of our 30 GHz data with the integrated flux at 1.5, 4.7 and 8.2 GHz \citep{car97}, Figure \ref{fig:index} shows that MRC~0943-242 has a steep spectral index between 1.5 GHz and 30 GHz with $\alpha = -1.44$. Similar to the case of MRC~2104-242, there is no evidence for spectral curvature within this range of frequencies.

No CO is detected at the central (nuclear) location of MRC~0943-242. When estimating an upper limit on $L'_{\rm CO}$ in MRC~0943-242 (potential 3$\sigma$ detection smoothed across 500 \kms), we derive $L'_{\rm CO} < 7.3 \times 10^{10}\ {\rm K~km~s^{-1}~pc^2}$ \citep[for $D_{\rm L} = 24242$ Mpc, which corresponds to a angular-size scale of 7.65 kpc/arsec for MRC~0943-242;][]{wri06}.

\subsubsection{Tentative off-nuclear CO detection}
\label{sec:tentative}

As can be seen in Fig. \ref{fig:MRC0943}{\sl b,d,e,} we find a tentative, off-nuclear 3$\sigma$ CO(1-0) detection in the Hanning smoothed data of MRC~0943-242 (with $\sigma$ the noise level at the frequency that corresponds to the tentative detection, see the arrow in Fig. \ref{fig:MRC0943}{\sl c}). The tentative CO signal spreads over an area about the size of one synthesised beam roughly 60 kpc NE of the centre of the host galaxy. It peaks at $v \sim -100$ \kms\ with a flux density of $\sim$\,1.8 \mjybm (with a tentative second peak present at the 1 \mjybm\ level around $v \sim -500$ \kms). The estimated luminosity of the tentative double-peaked CO signal is $L'_{\rm CO} \sim 8 \times 10^{10}\ {\rm K~km~s^{-1}~pc^2}$ (Equation \ref{eq:lco}). Both the H75 and H168 array data show indications for this tentative CO signal (Fig. \ref{fig:MRC0943}{\sl e}). However, because of the low-level (1$\sigma$) structure in the noise/bandpass beyond the nominal 7mm observing band (see Sect. \ref{sec:CABBperformance}), our results did not improve by further smoothing/binning the data in velocity. Our tentative 3$\sigma$ detection thus needs to be verified with additional observations before conclusions can be drawn.

\subsection{The environments of high-$z$ radio galaxies}
\label{sec:stacking}

The large instantaneous velocity coverage of CABB (see Sect. \ref{sec:observations}) also makes it possible to search for CO emitters in the field of our high-$z$ radio galaxies. The full width half maximum (FWHM) of the primary beam is 87\,/\,95 arcsec at 33\,/\,30 GHz, corresponding to about 0.69\,/\,0.73 Mpc at the redshift of MRC~2104-242\,/\,MRC~0943-242.

\begin{figure}
\centering
\includegraphics[width=0.47\textwidth]{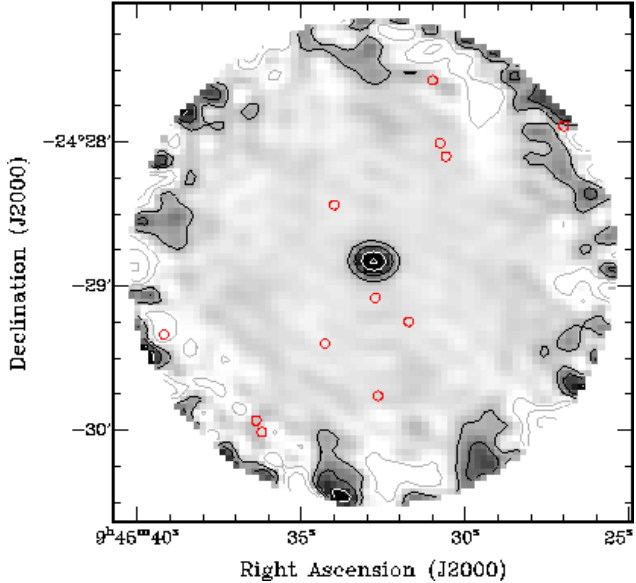}
\caption{Continuum image of MRC~0943-242. Shown is an area the size of the FWHM of the ATCA's primary beam at 30 GHz. Flux densities are corrected for the primary beam response pattern, which causes the effective noise level to rise significantly towards the edge of the image. Contours: -0.4, -1.0, -2.0 (grey); 0.4, 1.0 2.0, 3.0 (black/white) \mjybm. The red circles indicate the 12 \lya-bright galaxies in the field of MRC~0943-242 (and with redshifts within the CABB observing band) detected by \citet{breukelen05} and \citet{ven07}.}
\label{fig:stack}
\end{figure}

High-$z$ radio galaxies are generally located in proto-cluster environments \citep[e.g.][]{pen00,ven07}. MRC~0943-242 is known to be located in a proto-cluster with many nearby companions detected in Ly$\alpha$ and with known redshifts \citep{ven07}. There are 12 known \lya\ companions within the primary beam and observing band of our observations \citep[][see also Fig. \ref{fig:stack}]{breukelen05,ven07}. None of these galaxies shows a clear CO detection above a 3$\sigma$ limit, after correcting for primary beam attenuation.

\lya\ observations of the environment of MRC~2104-242 are lacking and hence the cluster properties are unknown. No CO was detected within the primary beam above 3$\sigma$ after correcting for primary beam attennuation. 

\section{Discussion}
\label{sec:discussion}

\subsection{H$_{2}$ masses}
\label{sec:h2masses}

CO is an excellent tracer of molecular hydrogen, because the rotational transitions of CO are excited primarily by collisions with H$_{2}$. A standard conversion factor $\alpha_{\rm x} = {\rm M}_{\rm H2}/{\rm L}'_{\rm CO}$ [M$_{\odot}$ (K \kms\ pc$^{2}$)$^{-1}$] is generally used to calculate the mass of the cold molecular gas \citep[where M$_{\rm H2}$ includes a fraction of the molecular gas that is in the form of helium -- see for example][for a review]{sol05}. For ultra-luminous infra-red galaxies (ULIRGs), \citet{dow98} derived a conversion factor of $\alpha_{\rm x} \sim 0.8$ M$_{\odot}$ (K \kms\ pc$^{2}$)$^{-1}$. This is in agreement with other observations of ULIRGs \citep{sol97,eva02} as well as high-$z$ sub-mm and star-forming galaxies \citep{tac08,sta08}, which imply that $\alpha_{\rm x} \sim 0.8 - 1.6$ M$_{\odot}$ (K \kms\ pc$^{2}$)$^{-1}$. We adopt a value of $\alpha_{\rm x} = 0.8$ M$_{\odot}$ (K \kms\ pc$^{2}$)$^{-1}$ also for the two high-$z$ radio galaxies that we study in this paper. We note, however, that there is a significant uncertainty in this conversion factor, since values as high as $\alpha_{\rm x} \sim 5$ have been derived for molecular clouds in the Milky Way \citep[][see also \citet{dic78}, \citet{blo86}, \citet{sol87}]{sco87,str88} as well as other nearby spiral galaxies \citep[][]{dic86,sol91}.

Based on our 3$\sigma$ upper limits on ${L}'_{\rm CO}$ and assuming $\alpha_{\rm x} = 0.8$, we estimate that M$_{\rm H2} < 2 \times 10^{10}$\,M$_{\odot}$ for MRC~2104-242 and M$_{\rm H2} < 6 \times 10^{10}$\,M$_{\odot}$ for MRC~0943-242. The tentative off-nuclear CO detection in MRC~0943-242 has an estimated molecular gas mass of M$_{\rm H2} = 6 \times 10^{10}$\,M$_{\odot}$.

\subsection{Molecular gas properties of high-$z$ radio galaxies}
\label{sec:comparison}

The upper H$_{2}$ mass limits that we derive for MRC~2104-242 and MRC~0943-242 are comparable to H$_{2}$ masses derived from CO detections in high-$z$ radio galaxies \citep[e.g.][see also \citet{sol05,mil08} for reviews]{sco97,pap00,bre03,bre03AR,bre05,kla05,nes09}. However, as discussed in Sect. \ref{sec:intro}, most of these observations have targeted the higher rotational CO transitions, which could underestimate the total molecular gas content in these systems. CO(1-0) detections have been claimed for two high-$z$ radio galaxies, namely 4C~60.07 \citep[$z=3.8$][]{gre04} and TNJ\,0924-2201 \citep[$z=5.2$][]{kla05}, both with M$_{\rm H2} = 1 \times 10^{11}$\,M$_{\odot}$. Our derived upper limit on molecular gas mass in MRC~2104-242 ($z=2.491$) is a factor 5 lower than this.

Sub millimetre galaxies (SMGs) are likely merging systems with a short-lived burst of extreme star formation and are believed to be the progenitors of local massive ellipticals \citep[e.g.][]{gre05,tac08}. In this sense, high-$z$ radio galaxies and SMGs could be the same type of objects that differ only in their level of AGN activity \citep[e.g.][]{reu07}, although \citet{ivi08} argue that the violent AGN activity may occur predominantly during the early evolutionary stages of these systems. \citet{gre05} derived a median cold gas mass of $\langle M_{\rm H2} \rangle = 3.0 \times 10^{10} M_{\odot}$ among 12 SMGs detected in CO \citep[see also][]{ner03}. This is of the same order as the upper limits that we derive for the mass of cold gas in MRC~2104-242 and MRC~0943-242.

Our derived upper limits on the molecular gas mass of MRC~2104-242 and MRC~0943-242 are {\sl lower} than the H$_{2}$ mass estimates for a non-negligible fraction of normal massive star forming galaxies at $z \sim 1-2$, derived from CO(3-2) observations by \citet[][even when accounting for the much larger CO-to-H$_{2}$ conversion factor that they used]{tac10}. A similar result is seen by comparing the upper limits on CO in samples of high-$z$ radio galaxies \citep{eva96,oji97} with the results of \citet{tac10}. Confirmation by observations of larger samples in the same CO transitions might indicate important differences in molecular gas fraction, excitation properties or chemical enrichment processes between high-$z$ radio galaxies and distant massive star forming galaxies.

The H$_{2}$ mass limit of MRC~2104-242 is only a factor 3 higher than the H$_{2}$ content of the most CO-bright radio galaxies in the low redshift Universe, as studied from CO(1-0) observatons of a large sample of IR-bright radio galaxies by \citet[][corrected for the difference in the used $\alpha_{\rm x}$-value and cosmological parameters]{eva05}. The vast majority of the low-$z$ radio galaxies in the sample of \citet{eva05}, however, contain significantly less molecular gas. This was recently confirmed by \citet{oca10} with a large sample of low-$z$ radio galaxies not selected on IR-properties, for which they derive a median \Htwo\ mass of only M$_{\rm H2} = 2.2 \times 10^8$ M$_{\odot}$.

We note that many high-$z$ CO detections to date are case-studies of galaxies that were pre-selected based on their properties at other wavelengths, such as a large sub-mm dust content or high infra-red (IR) luminosity. Both at low- and high-$z$ there appears to be a relation between the far-IR (FIR) and CO luminosity in different types of galaxies \citep[see][and references therein]{eva05,gre05}. Such a relation would indicate that (radio) galaxies with a FIR luminosity in the range of ULIRGs ($L_{\rm FIR} > 10^{12} L_{\odot}$) contain a CO luminosity similar to the upper limit that we derive for MRC~2104-242 ($L'_{\rm CO} \sim {\rm few} \times 10^{10}$ K \kms\ pc$^{2}$). From Spitzer observations of MRC~0943-242 at 24, 70 and 160$\mu$m \citep{sey07}, we estimate an upper limit on the total IR luminosity of $L_{\rm TIR} < 2 \times 10^{13} L_{\odot}$ when using the approximation by \citet{dal02}. Following the IR-CO relation found by \citet{eva05} and \citet{gre05}, this IR limit corresponds to an average CO luminosity roughly a factor 2 lower than the $L'_{\rm CO}$ upper limit that we derive for MRC~0943-242. The lack of detectable amounts of CO gas in MRC~0943-242 is therefore not unusual based on its IR properties, but it shows that unbiased CO(1-0) observations of high-$z$ radio galaxies are becoming feasible.

{\sl Systematic searches for various CO transitions in unbiased samples of high-$z$ (radio) galaxies are necessary to objectively investigate the overall content of cold molecular gas in the Early Universe.} Our results show that systematic and reliable searches for the ground-transition of CO in high-$z$ (radio) galaxies are becoming feasible with existing broadband facilities that can target the 20-50 GHz regime, such as the ATCA and EVLA.

\subsubsection{CO in the vicinity of MRC~0943-242?}
\label{sec:COvicinity}

In this Section we briefly discuss the possible nature of the tentative CO detection in the vicinity of MRC~0943-242, which needs to be confirmed before a more detailed analysis is deemed suitable.

The tentative CO detection ($\sim 60$ kpc NE of the host galaxy) may be associated with a nearby companion galaxy, although no companion has been detected in \lya\ at that location \citep{breukelen05,ven07}, so any such galaxy would have to be \lya-faint. Alternatively, the tentative CO detection may represent cold gas in the outer part of the quiescent \lya\ halo \citep{vil03}. \citet{bin00} show that \CIV\ absorption is associated with the deep \lya\ absorption in MRC~0943-242 and derive that this reservoir of absorbing gas is also located in the outer halo (i.e. ouside the radio cocoon). If confirmed, the cold gas properties of MRC~0943-242 resemble those found in the high-$z$ radio galaxies TXS\,0828+193 \citep[$z = 2.6$][]{nes09} and B3~J2330+3927 \citep[$z=3.1$][]{bre03}. 

The only two known high-$z$ radio galaxies in which CO(1-0) has been detected (4C\,60.07 and TNJ\,0924-2201; see Sect. \ref{sec:comparison}) also show indications that the CO gas may not be aligned with the central location of the host galaxy \citep[][]{kla04,ivi08}. In particular 4C\,60.07 shows an apparent deficit of molecular gas in the radio host galaxy, while CO appears to be present in a merging companion and associated tidal debris \citep{ivi08}. If confirmed, a more detailed comparison between the CO(1-0) properties of these systems deserves further attention.

The position angle of the radio source in MRC~0943-242 (which has a total linear size of about 4$^{\prime\prime}$ and is therefore unresolved in our observations) is PA = -74$^{\circ}$ \citep{car97}. This is roughly within 45$^{\circ}$ of the location of the tentative CO detection from the central region of the radio host galaxy. If confirmed, this may resemble alignments that \citet{kla04} argue exist among other high-$z$ radio galaxies.

\subsection{Radio continuum}
\label{continuum}
 
Both radio sources are clearly detected in our sensitive ($\sigma \sim 30$ $\mu$Jy) 7mm continuum observations. Their spectral indices are relatively steep from $\sim$30 GHz down to 1.4 GHz, with no evidence for spectral curvature within this large range of frequencies. This indicates that there is no turn-over due to synchrotron losses or inverse Compton cooling up to $\sim$115 GHz in the restframe of these radio sources. This is consistent with continuum observations of a large sample of high-$z$ ultra-steep spectrum radio galaxies by \citet{kla06}, who also find relatively steep power law spectral energy distributions (SEDs) with no evidence for spectral steepening up to several tens of GHz in the rest frame. A detailed analysis of this phenomenon is crucial for understanding the electron acceleration mechanism or environmental properties of high-$z$ radio sources, but is beyond the scope of this paper.

\section{Conclusions}
\label{sec:conclusions}

We presented the first 7mm observations of two high-$z$ radio galaxies (MRC~2104-242 and MRC~0943-242) with the $2 \times 2$~GHz Compact Array Broadband Backend. Our results demonstrate the feasibility of using ATCA/CABB for spectral-line work at high redshift. We also presented 7mm continuum images of the two high-$z$ radio galaxies, with a typical rms noise level of $\sim$30 $\mu$Jy beam$^{-1}$. The enhanced spectral-line and continuum capabilities of ATCA/CABB in the millimetre regime complement those of other large existing and upcoming observatories, such as PdbI, EVLA and ALMA.

From our CO(1-0) data we derive upper limits on the H$_{2}$ mass of M$_{\rm H2} < 2 \times 10^{10} M_{\odot}$ for MRC~2104-242 and M$_{\rm H2} < 6 \times 10^{10} M_{\odot}$ for MRC~0943-242 ($\alpha_{\rm x} = 0.8$). These upper limits are of the same order as H$_{2}$ mass estimates derived from CO detections of other high-$z$ radio galaxies and SMGs, but lower than the mass of molecular gas detected in a non-negligible fraction of normal star forming galaxies at $z \sim 1-2$. For MRC~0943-242 we also find a tentative CO(1-0) detection at about 60 kpc distance from the central region of the host galaxy, but this needs to be confirmed with additional observations.

The spectral index of both MRC~2104-242 and MRC~0943-242 is relatively steep with $\alpha \approx -1.5$ between 1.4 and 30 GHz. There is no evidence for spectral curvature up to $\sim 115$ GHz in the rest frame of these radio sources.

\section*{Acknowledgments}
We are tremendously grateful to Warwick Wilson, Dick Ferris and their team and to the engineers and system scientists in Narrabri for making CABB such a great success. We also thank the anonymous referee for good suggestions that significantly improved this paper. The Australia Telescope is funded by the Commonwealth of Australia for operation as a National Facility managed by CSIRO.


\end{document}